\documentclass [12pt]{article}
\usepackage{epsfig}
\begin{document}

\begin{center}
{\bf Quantum Hall Effect and Dyson-Swinger Equations }

\vspace{0.8cm} {P.A. Kurashvili}

\vspace{0.5cm}\baselineskip=14pt

\vspace{0.5cm} \baselineskip=14pt {\it Department of Theoretical
Physics, Tbilisi State University, 0128 Tbilisi, Georgia}

\vspace{0.5cm}
\begin{abstract}In this paper we make attempt to obtain a description of the
Quantum Hall Effect (both integer and fractional) by means of
electron's Green functions of three-dimensional (planar)
electrodynamics. We show that expression for the free electron
propagator yields an integer number for the second Chern-Simons
term, that corresponds to the quantized Hall conductivity in the
approximation of non-interacting particles for integer filling
factors, when there exists a gap for all excitations in the
system. Then we try to check correspondence between fractional
case and "interacting" Green functions, so it requires taking into
consideration "full-fledged" propagators, including interactions.
We are going to obtain
them from Dyson-Swinger equations. We attempt to reach out from the
perturbation theory regime using a specific method,
called scale approximation. Our solutions are found in general
gauge.

\end{abstract}

\end{center}

\section{Introduction}The Quantum Hall Effect , found more then twenty
years ago \cite{KDP}, \cite{TG},  have been a subject of great interest
since its
discovery. As there was discovered, the planar system of electrons, placed
in strong homogenous magnetic field perpendicular to the plane at low
temperatures
shows unusual behavior near the values of the field strengths
corresponding to the integer fillings of Landau levels. The
antisymmetric part of the conductivity tensor corresponding to the
Hall conductance appears constant for the certain ranges of values
of the magnetic field instead of equally increasing, as one could
suppose starting from naive microscopic considerations, while the
diagonal part is almost equal to zero at the same ranges. The
quantization value of the Hall conductance have been shown to be
equal to the ratio of two fundamental physical constants - the
elementary electric charge and the Plank constant $\frac{e^2}{h}$.

The explanation for this behavior have been found in extremely
"quantum" nature of electrons in two-dimensional systems due to
the strong magnetic field and particularities of the single electron
spectrum, when Landau levels, corresponding to the single-particle
spectrum get widened due to local impurities and defects and there
exist regions of localized states between them, interfering to the
further increase of the Hall conductance while the Fermi level
"travels" through these regions \cite{Pr}.

The precise quantization of the Hall
conductance have been explained in works of Laughlin \cite{Lau1}, Haldane
\cite{Hal},
Thouless et al. \cite{TKN} and others. The quantized Hall conductance
expressed as a topological invariant, corresponds to an integer
number in the case of non-degenerate ground (vacuum) state, or to a
simple fraction with an odd denominator when any degeneracy is
relevant. In works of Robert Laughlin there have been discovered
some unusual properties of Quantum Hall state excitations, such as
modes with fractional charges and statistics \cite{Lau2}, \cite{Lau3}. After
works of
Zhang, Hu and others the field-theory language descriptions of the
Quantum Hall Liquid was obtained that revealed the parallels to
superconductivity and three-dimensional $2+1$ Chern-Simons
Electrodynamics. According to these models the Quantum Hall conductance is
interpreted as a
Chern-Simons mass of the gauge field. In this paper
we follow the relativistic approach based on relativistic Kubo
formula for the quantized Hall conductance  and the Dyson-Swinger
equation for the full electronic propagator \cite{AN1}, \cite{AN2}.

\section{Derivation of Kubo formula}

The Hall conductivity is
defined as the linear part of the correlation function \cite{AN1}:
\begin{equation}\Pi_{\mu\nu}(q)=\frac{1}{(2\pi)^3}\int d^3e^{\imath
qx}\langle0|J_\mu(x)J_\nu(0)|0\rangle.\end{equation}

Where
$J_\mu(x)$ is the conserved current:\begin{equation}\partial^\mu
J_\mu(x)=0.\end{equation}
Inserting this equation
to the above expression we obtain:
$$\Pi_{\mu\nu}(q)=-\frac{e^2}{(2\pi)^3}\int
d^3Tr(\Lambda_\mu(p,p+q)S(p)\Lambda_\nu(p,q+p)S(p-q)).$$

We used the following definitions:\begin{equation}\int d^3x d^3x_1 d^3y_1
e^{\imath(p\prime{x}_1-py_1-qx)}\langle0|T(J_\mu(x)\psi(x_1)\psi^+(x_2))|0\rangle=$$$$S(p\prime)\Lambda_\mu(p\prime,q+p)S(p)(2\pi)^3\delta^{(3)}(p\prime-p-q),\end{equation}
and\begin{equation}\int d^3x_1
d^3y_1e^{\imath(p\prime{x}_1-py_1)}\langle0|T(\psi(x_1)\psi^+(y_1)|0\rangle.\end{equation}

The linear part of current correlation function is
given by $$\frac{\partial}{\partial
q^\alpha}\Pi_{\mu\nu}|_{q=0}=-\frac{e^2}{(2\pi)^3}\int d^3p
Tr\{[\frac{\partial}{\partial
q_\alpha}\Lambda_\nu(p,p+q)]_{q=0}S(p)\Lambda_\nu(p,p)$$$$+\Lambda_\mu(p,p)S(p)\partial
q_\alpha\Lambda_\nu(p,p+q)|_{q=0}S(p)-\Lambda_\mu(p,p)S(p)\Lambda_\nu(p,p)\frac{\partial}{\partial
p_\alpha}S(p)\}.$$

The first two terms give no contribution to the
antisymmetric part of the conductivity tensor, which can be
expressed as
\begin{equation}\sigma_{xy}=\lim_{\varepsilon\longrightarrow
0}\frac{\varepsilon_{\mu\nu\lambda}q^2\Pi_{\mu\nu}(q)}{q^2}.\end{equation}

And the antisymmetric part of this is given
by:$$\frac{1}{6}\varepsilon^{\mu\nu\rho}\int\frac{d^3p}{(2\pi)^3}(\frac{\partial
S^{-1}(p)}{\partial p_\mu}S(p)\frac{\partial S^{-1}(p)}{\partial
p_\nu}S(p)\frac{\partial S^{-1}}{\partial
p_\rho}S(p)).$$

Finally we arrive at
\begin{equation}N=\frac{1}{24\pi^2}\int
d^3p\epsilon^{\mu\nu\rho}Tr\{(\partial_\mu S^{-1})S(\partial_\nu
S^{-1})S(\partial_\rho S^{-1})S(p)\}.\end{equation}

Deriving this expression we made use of the Ward-Takahashi identity
$\Lambda_\mu(p,p)=\frac{\partial S^{-1}(p)}{\partial
p^\mu}$.
$N$ denotes the topological invariant of mapping from
3-dimensional momentum $p$-space to matrices $S(p)$ as if the
integration area were 3-dimensional sphere $S^3$. When the value
of momentum tends to infinity, the direction plays no role in our
consideration
and one can stereographically map
$R^3$ to $S^3$. $N$ topological invariant known as Pontriagin
index and is always integer. Therefore the antisymmetric part of the
conductivity tensor is given by the expression:
\begin{equation}\sigma_{xy}=\frac{e^2}{2\pi\hbar}N=\frac{e^2}{h}N,\end{equation}
where $N$
is some integer. So far, we have seen that Hall constant is quantized
in the framework of the relativistic field theory. Below we  attempt
to get those numbers from the microscopical theory considerations.

First let us assume that fermions in the loop are non-interacting, so we can
insert free propagator $S^{-1}=\hat{p}-m$ directly into our formula and
obtain:
\begin{equation}N=\frac{\imath}{2\pi^2}\int
d^3p\frac{m(p^2-m^2)}{(p^2-m^2)^3}=-\frac{1}{2}.\end{equation}

After the integration
we must multiply this result by two, because we have two spin
degrees of freedom for an electron and finally get an integer number: $N=1$.

\section{Dyson-Swinger equation}

Dyson-Swinger equation is one of the most sensible ways of finding
non-perturbational solutions
in the quantum field theory. As
we have seen already, the quantum Hall conductivity can be expressed by
the integral that contains the electron propagator $S(p)$. The
computation of this quantity is very difficult task and is the
object of studies up to the present days. The main difficulty is
that the Dyson-Swinger system is not closed and self-contained and one have
to make additional assumptions to arrive at the closed and
self-consistent system of equations.

For simplicity let us write
our equations for the simplest case of the zero bare mass (in what follows
we
also neglect the Chern-Simons mass of photon):
\begin{equation} S^{-1}=\hat{p}-\imath\frac{e^2}{(2\pi)^3}\int
d^3k\gamma^\mu D_{\mu\nu}(p,k)S(k)\Gamma^\nu(p,k)\end{equation}

On
the right-hand side of this expression there stands the electron-photon
vortex function, $\Gamma^\nu(p,k)$ and the full propagator of photon
$D_{\mu\nu}(p-k)$. One can write down the Dyson-Swinger equation
in the graphical form.

%

\includegraphics[bb = 34 583 578 707,width=.8\textwidth]{fig3.eps}

We can see, that right-hand side contains a new
factor $K$, electron-electron scattering amplitude. In order to
obtain a closed system, one has to make any assumption for vertex
function an for $D_{\mu\nu}$, only then can this be
considered as an equation.

One of the most convenient ways of studying the
equation at hand is so-called ladder approximation, when one
replaces the full vortex function by simple factor
$\gamma_\mu$ and the full photon propagator is substituted by free
one. This approximation was widely used in the middle of the last century
by Landau, Berestetsky an Pitaevski during studies of the bound systems. It
inherits many features of
the perturbation theory but also includes summation over diagrams
of certain type as a way of reaching out of the perturbation
approximation, reliable only in the week coupling regime.

The second
method mostly used at last times is based on approximation that
conserves gauge invariance at every succeeding step of
calculations. The method makes wide use of Ward-Takahashi
identities conserving gauge invariance at every step of
calculations. By this method one can explicitly find the
longitudinal part of the vortex function but the transversal part
remains totally undefined. This method is frequently used in studies
of the infrared part of electron propagator. In our case this
method is of no use, despite its non-pertubavity, because all
solutions found in literature are constructed so that vortex
factor is proportional to the transferred momentum, that has to
be multiplied by photon propagator and gives zero because of
gauge invariance. In other words, the longitudinal factor falls
out from the Dyson-Swinger equation and the information is lost. The
transversal part is completely undefined and this method is
useless for our task. Therefore one can have more profit using the ladder
approximation.

\section{Landau gauge} We begin with the Landau
gauge when the photon propagator is (we assume zero Chern-Simons mass)
\begin{equation}D_{\mu\nu}=-\frac{1}{q^2}(g_{\mu\nu}-\frac{q_\mu
q\nu}{q^2})\end{equation}

The electron propagator is written in form:
\begin{equation}S^{-1}(p)=\beta(p^2)\hat{p}-\alpha(p^2)\end{equation} Where
$\alpha$ and $\beta$ are functions, we have to find. Now let us put
this expression into DS equation and use the properties of
three-dimensional Dirac matrices:
$$Tr\gamma^\mu=0,$$ $$Tr\gamma^\mu
\gamma^\nu=2g_{\mu\nu},$$
$$Tr\gamma^\mu\gamma^\mu\gamma^\lambda=-2\imath\epsilon^{\mu\nu\lambda},$$
$$Tr\gamma^\mu\gamma^\nu\gamma^\lambda\gamma^\sigma=2[g^{\mu\nu}g^{\lambda\sigma}+g^{\mu\sigma}g^{\nu\lambda}-g^{\mu\lambda}g^{\nu\sigma}],$$
$$Tr\gamma^\mu\gamma^\nu\gamma^\lambda\gamma^\sigma\gamma^\rho=-2\imath[g^{\mu\nu}\epsilon^{\lambda\sigma\rho}+g^{\lambda\sigma}\epsilon{\mu\nu\rho}+g^{\sigma\rho}\epsilon{\mu\nu\lambda}-g^{\lambda\rho}].$$

We also employ following relations:
$\epsilon^{\mu\nu\sigma}g_{\sigma\lambda}=\epsilon_\lambda^{\mu\nu}$
Then we can derive the properties of the $\gamma$ matrices of the
form
$\gamma^\mu\gamma^\nu=g^{\mu\nu}-\imath\epsilon^{\mu\nu\lambda}\gamma_\lambda$

After calculating traces from both
sides
$$\hat{p}(\beta(p^2)-1)-\alpha(p^2)=\frac{\imath
e^2}{(2\pi)^3}\int
d^3\frac{[2\alpha(k^2)-2\beta(k^2)\frac{\hat{k}(q,k)}{q^2}]}{q^2[(k^2)\beta^2(k^2)-\alpha^2(k^2)]}$$

Then let us multiply both sides of the expression by $\hat{p}$ and
again take the
trace:$$\alpha(k^2)=-2\imath\frac{e^2}{(2\pi)^3}\int
d^3k\frac{\alpha(k^2)}{q^2[k^2\beta(k^2)-\alpha(k^2)]}$$

If we integrate by angles at right-hand side of this expression we get zero
and
therefore.
$$\beta(p^2)-1=-2\imath\frac{e^2}{(2\pi)^3}\int
d^3k\frac{\beta(k^2)}{k^2\beta^2(k^2)-\alpha(k^2)}\frac{(q,k)(p,q)}{q^4p^2}$$

Calculating traces of gamma-matrices we get the
closed expression for $\alpha$:
\begin{equation}\alpha(k^2)=-2\imath\frac{e^2}{(2\pi)^3}\int
d^3k\frac{\alpha(k^2)}{q^2[k^2-\alpha(k^2)]}\end{equation}

This
equation is strongly nonlinear and one can not resolve it
explicitly, therefore we need to make any reasonable
approximation for it. The tool for performing this item have been
provided by Maris, Hertscovitz and Jacob \cite{MHJ}, they showed that an
approximation $\alpha\equiv m$ is indeed very powerful
non-perturbative supposition in many problems, when the gap function
$\alpha$ is replaced
by the physical mass of electron: We can express the self-energy
(in Euclidian momenta) so: $\alpha(-p^2)=(p^2+m^2)\chi(p^2)$ and
perform the Furrier-transform to new variables:
$$\chi(r)=\frac{1}{(2\pi)^3}\int
d^3k\chi(k^2)e^{\imath\overrightarrow{k}\overrightarrow{r}}$$
Then we obtain the differential equation:
\begin{equation}\frac{d^2}{dr^2}\chi(r)+\frac{2}{r}\frac{d}{dr}\chi(r)-(m^2-\frac{e^2}{2\pi
r})\chi(r)=0\end{equation}

This expression is very similar to one
for the wave function of electron in zero angular momentum state
of hydrogen atom. The solution of this equation has the form
of hypergeometrical function $$\chi(r)=e^{\nu
r}F(a,c,\frac{x}{\lambda})$$ Here we must take $\nu=m$ and assume
that $a$ is equal to any negative integer $-n$ or to zero
in order to obtain a solution, falling exponent mutilplied by a
finite polinomial, that decreases to zero on spatial infinity:
$a=1-\frac{e^2}{4\pi m}$hence, we conclude, that the "mass" of
electron is $m=\frac{e^2}{4\pi(n+1)}$, resembling the expression
for the energy levels of the hydrogen atom. In the simplest case
of $n=0$ we have: $$\chi(r)=Ce^{-mr}.$$

We convert this expression into the momentum representation again
and choose the value of the arbitrary constant $C$ in order to get
$\alpha(p^2=0)=m.$  From this by Furrier-transformation we get for
$\alpha$ in momentum representation:
$$\alpha(-p^2)=\frac{m^3}{p^2+m^2}.$$

If we insert
again this expression to our formula, then
\begin{equation}N=-\frac{2}{3\pi}\int
p^2dp\frac{3\alpha(-p^2)-2p^2\alpha'(-p^2)}{(p^2+\alpha^2(-p^2))^2}\end{equation}

Thus $$N=-\frac{2}{3\pi}\int
p^2dp\frac{m^3(5p^2+3m^2)}{(p^2+m^2)^4}$$

This
integral can be rewritten as $$N=-\frac{1}{3\pi}\int
dx\frac{\sqrt{x}(x+5x}{(1+x)^4}$$

and the integration
yields the result in terms of the beta function:
\begin{equation}N=-\frac{8}{3}B(\frac{3}{2},\frac{5}{2})\end{equation}.

Finally we obtain:
\begin{equation}N=-\frac{1}{6}\end{equation}.

After taking into account the
multiplicity due to the electron spin states, we obtain the result
$$\sigma_{xy}=\frac{1}{3}\frac{e^2}{2\pi}$$

\section{Solving of DS equations}
In the previous section we obtained the quantized Hall conductance
by means of Kubo's relativistic formula and expression for the
free electron propagator. As we mentioned above, in the framework
of non-relativistic theory one has to take into account the
interactions between the charge carriers (electrons) in order to
obtain fractional numbers. On relativistic language this imports
that one have to insert the full propagator obtained from
Dyson-Swinger equations.

In previous section we made it for Landau
gauge and obtained a simple fraction, now we are going to solve
the problem in the general gauge and see, if it is possible to
obtain any other simple fractions and the explicit form for
electron propagator. In general gauge, the propagator for the
free photon is:
\begin{equation}D_{\mu\nu}(k)=\frac{1}{k^2}[(g_{\mu\nu}-\frac{k_\mu
k_\nu}{k^2})+a\frac{k_\mu k_\nu}{k^2}\end{equation}

If we denote
the vortex function by $\Gamma_\nu(p,k)$ , it is possible to
decompose it into two, longitudinal and transverse parts:
\begin{equation}\Gamma_\nu(p,k)=\Gamma_\nu^T(p,k)+\Gamma_\nu^L(p,k)\end{equation}

Here
$$\Gamma_\nu^L(p,k)=\frac{q_\nu}{q^2}[S^{-1}(p)-S^{-1}(k)]$$
$q=p-k$ and $$\Gamma_\nu^T=(g_\nu\rho-\frac{q_\nu
q_\rho}{q^2})\Gamma^\rho(p,k)=g_\nu\rho-\frac{q_\nu
q_\rho}{q^2})\gamma^\rho\beta(p^2_{max})$$

The first of them is
taken in order to satisfy Ward-Takahashi identities. In Landau
gauge this part disappears from the equation.
After inserting this expression the initial equation gets the following
form:
\begin{equation}-\alpha(p^2)+\hat{p}\beta(p^2)=
\hat{p}-\frac{\imath e^2}{(2\pi)^3}\int
d^3k\frac{1}{q^2[k^2\beta(k^2)-\alpha(k^2)]}$$$$[(2\alpha(k^2)-\beta(k^2)\hat{q}\frac{(k,q)}{q^2})\beta(p^2_{max})+a\hat{q}(\hat{k}\beta(k^2)+\alpha(k^2))(\hat{p}\beta(p^2)-\alpha(p^2))]\end{equation}

Here we again use the properties of three-dimensional Dirac
matrices and compute the traces from both sides of the equation,
then multiplying the equation by $\hat{p}$ to obtain:
\begin{equation}\alpha(-p^2)=-\frac{e^2}{(2\pi)^3}\int
d^3\frac{1}{q^2[k^2\beta^2(-k^2)+\alpha^2(-k^2)]}$$$$[2\alpha(-k^2)\beta(-p^2_{max})-\frac{2a}{q^2}[\alpha(-k^2)\beta(-p^2)(p,q)-\alpha(-p^2)\beta(-k^2)(q,k)]]\end{equation}

\begin{equation}p^2(\beta(-p^2)-1)=\frac{e^2}{(2\pi)^3}\int
d^3k\frac{1}{q^2[k^2\beta^2(-k^2)+\alpha^2(-k^2)]}$$$$\frac{2a}{q^2}[\beta(-k^2)\beta(-p^2)p^2(q,k)+\alpha(-k^2)\alpha(-p^2)(p,q)]\end{equation}

In Landau gauge we had $\beta\equiv1$. This approximation would
enable us to simplify the first expression. In order to clarify the
reliability of
this approximation, we have to evaluate the $\alpha$. We work in
linear approximation, where $\alpha$ is so small, that one may
neglect its value in the denominator under integral. Then we get
the closed expression for $\beta$:
$$p^2(\beta(-p^2)-1)=2a\frac{e^2}{(2\pi)^3}\int
d^3k\frac{p^2(p,k)}{q^4k^2\beta^2(-k^2)}\beta(-k^2)\beta(-p^2)$$

or
$$1-\frac{1}{\beta(-p^2)}=2a\frac{e^2}{(2\pi)^3}\int
d^3k\frac{p^2(q,k)}{q^4k^2}\frac{1}{\beta^2(-k^2)}$$

The solution of this equation is possible by Furrier-transform
method: if we put $\frac{1}{\beta(-p^2)}=p^2\varphi(p^2)$ and
transform it to new variables: $\varphi(p^2)=\int
d^3e^{\imath\overrightarrow{p}\overrightarrow{r}}\varphi(r)$, we
arrive at the following equation for the radial function:
$$r^2\frac{d^2\varphi}{dr^2}+(2r-2abr^2)\frac{d\varphi}{dr}-abr\varphi(r)=\frac{1}{4\pi}\delta(r)$$

Here $b=\frac{e^2}{2\pi}$. If we denote $abr=\rho$, we get:
$$\rho^2\frac{d^2\varphi}{dr^2}+2\rho(1-\rho)\frac{d\varphi}{dr}-\rho\varphi=\frac{ab}{4\pi}\delta(\rho)$$

Because of the presence of the singular Dirac function in the
right side, we have to perform one more Furrier transformation to
variables $k$: $\varphi(k)=\int d\rho\varphi(\rho)e^{\imath
k\rho}$ Then after the denomination $\imath k=t, \varphi'(t)=p(t)$
$$\varphi''(t)(t^2+2t)+(2t+5)\varphi'(t)=\frac{\lambda}{4\pi}$$

From this non-homogenous equation we find:
$$p(t)=\varphi'(t)=\frac{1}{\pi}(\frac{1}{2}\frac{\lambda}{t}-\frac{3}{4}\frac{\lambda}{t^2}-\frac{3}{2}\lambda\frac{1}{t^{5/2}(t+2)})\ln\frac{s-1}{s+1}$$

Where $\lambda=ab$, $s=(\frac{t+2}{t})^{\frac{1}{2}}$ We can
reconstruct $\varphi(p^2)$ by reversing Furrier integrations:
$$\varphi(p^2)=\int
d^3re^{\imath\overrightarrow{p}\overrightarrow{r}}\varphi(r)=\frac{\pi}{\imath\rho}\int
\varphi(r)(e^(\imath pr)-e^{-\imath
pr})rdr=\frac{\pi}{\imath\rho}\int\frac{dk}{2\pi}\int
dr\varphi(k)e^{-\imath k\lambda r}(e^{\imath pr}-e^{e-\imath
pr})r=$$$$\frac{\pi}{p}\frac{\partial}{\partial p}(\int
dk\varphi(k)\delta(k\lambda+p)+\int
dk\varphi(k)\delta(k\lambda-p))=$$$$\frac{\pi}{p}(\varphi'(k)-\varphi(-q))=\frac{1}{p^2}+\frac{3}{8}\frac{\lambda^2}{p^2}[\frac{\lambda}{(\imath
p)^{1/2}(\imath p + 2 \lambda)^{1/2}}\ln
\frac{s-1}{s+1}-\frac{\lambda}{(-\imath p)^{1/2}(-\imath
p+2\lambda)^{1/2}}\ln\frac{s^*-1}{s^*+1}$$

$\beta(p^2)$ is
equal to unity when $p=0$ and $p=\infty$ and very close to this
value at intermediate values of the argument and in the case of
relevance we can put it equal exactly to one and solve the
equations for $\alpha$:
\begin{equation}\alpha(-p^2)=\frac{e^2}{(2\pi)^3}\int\frac{d^3k}{q^2(k^2+m^2)}\{2\alpha(-k^2)-\frac{2a}{q^2}[\alpha(-k^2)(p,q)-\alpha(-p^2)(q,k)]\}\end{equation}

Here again we use substitution $\alpha(-p^2)=(p^2+m^2)\chi(p^2)$
and Furrier-transformation to real variables.

First we get:
$$(p^2+m^2)[1-\frac{e^2a}{(2\pi)^3}(\frac{2\pi^2}{p}\arctan\frac{m}{p}-2\pi^2\frac{m}{p^2+m^2})]\chi(p^2)=$$$$\frac{2e^2}{(2\pi)^3}\int\frac{d^3k}{q^2}(1-\frac{a}{q^2}(p,k))\chi(k^2)$$

The arctan on the left-hand side is suggestive of very strong
non-linearity. We may substitute it by a $\frac{\pi}{4}$ if we put
that space part of the momentum is small and the value of the total
momentum is the of the same order as mass.

After this approximations and
transformation to radial variables, we have:
\begin{equation}r\frac{\partial^2}{\partial r^2}\chi(r)+(2-\frac{e^2a}{4\pi
B}r)\frac{\partial}{\partial r}\chi(r)-[(m^2-\frac{e^2a}{4\pi
B})r-\frac{e^2}{2\pi B}]\chi(r)=0\end{equation} Here
$B=1-\frac{e^2a}{16\pi}$

This equation can be rewritten as one for the for the degenerate
hypergeometrical function and its solution is: $$\chi(r)=e^{\nu
x}F(a,2,\frac{x}{\lambda})$$

Where $a=-2\nu-\frac{e^2}{2\pi B}$. It can be only zero or any negative
integer value to fall in infinity. The
quantities $\nu$, $\lambda$ are defined from the following
equations: $$\nu^2-\frac{e^2a}{4\pi B}-(m^2-\frac{e^2a}{4\pi B}),$$,
$$0=1+2\lambda(2\nu-\frac{e^2a}{4\pi B})$$

Going to the previous representation again,
we can extract the propagators in momentum representation again.  One also
can see from this solutions, that in Landau gauge, this expression coincides
with the one, obtained in the previous section.

\section{Concluding remarks}
In this paper we have outlined the scheme of calculating the topological
Quantum Hall numbers using the relativistic microscopical theory.

Besides some
successes, one can pose some questions, that are difficult to answer.
First, how one can obtain all Quantum Hall fractures, basing on this theory?
The calculations of the filling factor based on further hypergeometrical
solutions
of equation in Landau gauge yielded very poor results,
compared with the brilliant $\frac{1}{3}$
for the case $n=0$ and the very sense of the "higher energy levels"
is still unclear. Also we had
very poor progress when trying to solve Dyson-Swinger equations in the
general gauge, because
removing the strong non-linearity gives only expression, trivially going to
the one obtained
in Landau gauge when gauge parameter $a$ tends to zero.

One can see, that this nonlinearity
needs more subtle treatment in order to maintain all
the information about the full propagator,
probably a slight modification of the ladder aproximation
itself, or considering strong-coupling
limit (corresponding to the big $n$-s in our solutions).

It would be of interest to try this
method to the new-found four and eight-dimensional
generalizations of the quantum Hall effect \cite{ZH}.
We shall try to do it in the following papers.

\section{acknowledgments}
I feel very grateful to my supervisor, prof. Anzor khelashvili
for introducing me to the the of Quantum Hall Effect,
also to Michael Maziashvili, Merab Gogberashvili and  Vasil Lomadze
for many helpful discussions on this subject and great
encouraging and support during this work.

\end{document}